\documentclass[aip,apl,preprint,showpacs,floatfix,superscriptaddress]{revtex4}
\usepackage{graphicx,amsmath,amssymb,ulem,color}

\newcommand{\be}{\begin{equation}}
\newcommand{\ee}{\end{equation}}
\newcommand{\bea}{\vspace{0.25cm}\begin{eqnarray}}
\newcommand{\eea}{\end{eqnarray}}

\begin{document}


\title{An extremely low-noise heralded single-photon source: a breakthrough for quantum technologies}

\author{G. Brida}
\affiliation{Istituto Nazionale di Ricerca Metrologica - I.N.RI.M., Strada delle Cacce 91, 10135
Torino, Italy}

\author{I. P. Degiovanni}
\affiliation{Istituto Nazionale di Ricerca Metrologica - I.N.RI.M., Strada delle Cacce 91, 10135
Torino, Italy}

\author{ M. Genovese}
\affiliation{Istituto Nazionale di Ricerca Metrologica - I.N.RI.M., Strada delle Cacce 91, 10135
Torino, Italy}

\author{  F. Piacentini }
\affiliation{Istituto Nazionale di Ricerca Metrologica - I.N.RI.M., Strada delle Cacce 91, 10135
Torino, Italy}

\author{ P. Traina}
\affiliation{Istituto Nazionale di Ricerca Metrologica - I.N.RI.M., Strada delle Cacce 91, 10135
Torino, Italy}

\author{A. Della Frera}
\affiliation{Politecnico di Milano, Piazza Leonardo da Vinci 32, 20133 Milano, Italy}

\author{A. Tosi}
\affiliation{Politecnico di Milano, Piazza Leonardo da Vinci 32, 20133 Milano, Italy}

\author{ A. Bahgat Shehata}
\affiliation{Politecnico di Milano, Piazza Leonardo da Vinci 32, 20133 Milano, Italy}

\author{ C. Scarcella}
\affiliation{Politecnico di Milano, Piazza Leonardo da Vinci 32, 20133 Milano, Italy}

\author{ A. Gulinatti}
\affiliation{Politecnico di Milano, Piazza Leonardo da Vinci 32, 20133 Milano, Italy}

\author{ M. Ghioni}
\affiliation{Politecnico di Milano, Piazza Leonardo da Vinci 32, 20133 Milano, Italy}

\author{S. V. Polyakov}
\affiliation{Joint Quantum Institute, University of Maryland, and National Institute of Standards and Technology, 100
Bureau Dr, Stop 8410, Gaithersburg, MD, 20899 USA}

\author{A. Migdall }
\affiliation{Joint Quantum Institute, University of Maryland,  and National Institute of Standards and Technology, 100
Bureau Dr, Stop 8410, Gaithersburg, MD, 20899 USA}

\author{A. Giudice}
\affiliation{Micro Photon Devices Srl, Via Stradivari 4, 39100 Bolzano, Italy}


\begin{abstract}

Low noise single-photon sources are a critical element for quantum technologies. We present a
heralded single-photon source with an extremely low level of residual background photons, by
implementing low-jitter detectors and electronics and a fast custom-made pulse generator
controlling an optical shutter (a LiNbO$_3$ waveguide optical switch) on the output of the source.
This source has a second-order autocorrelation $g^{(2)}(0)=0.005(7)$, and an \textit{Output Noise
Factor} (defined as the ratio of the number of noise photons to  total photons at the source output
channel) of $0.25(1)\%$. These are the best performance characteristics reported to date.

\end{abstract}

\maketitle

Single-photon sources play a key role in many applications such as quantum metrology
\cite{metrology_1,metrology_2,metrology_3,metrology_4,metrology_5}, quantum information
\cite{quantumcomm,quantumtech}, and in the testing of the foundations  of quantum mechanics
\cite{alicki_1,alicki_2,alicki_3}, so there is much interest and effort in improving the
performance of such sources \cite{shell_2, alan}.

An ideal single photon source would be one for which: a single photon can be emitted at any arbitrary time defined
by the user, the probability of a single-photon emission  at that time is 100$\%$, and the
probability of multiple-photon emission is 0$\%$, subsequent photons are indistinguishable, and the
repetition rate is arbitrarily fast \cite{shell_2, alan}. Such a source would truly be a
``deterministic on-demand'' single-photon source.

Unfortunately, deviations from these ideal characteristics are always present in real-world
sources, and must be considered when using them in specific applications. The effect and emergence
of this issue is seen in the  developmental history of single-photon sources, where while initial
efforts focused mostly on increasing rates of photon generation and the efficiency of that
generation, current improvement efforts are more driven by the requirements of particular
applications. As a result, efforts now often deal with improving more than one characteristic
simultaneously, as it is well understood that heroic results in improving just a single parameter
are often of little practical use \cite{shell_2, alan}.


To achieve fully deterministic single-photon generation it is natural to employ sources whose
physics guarantees deterministic single-photon emission - a key advantage over sources that rely on
inherently probabilistic generation. Such deterministic sources could be single quantum emitters:
single atoms, ions, molecules, quantum dots, or color centers in diamond. But in practice the
distinction between deterministic and probabilistic sources often blurs due to issues such as non
unity extraction efficiency, a probabilistic loss that must be convolved with the deterministic
generation to get the actual performance of a source. As extraction loss increases, a theoretically
deterministic source becomes more probabilistic in operation. This makes it clear that improving
only one characteristic of one portion of the source (e.g. probabilistic vs. deterministic
character of photon generation) is insufficient, and that this characteristic of the overall source
is really a continuum \cite{shell_2, alan}.

One performance measure important for many applications and one that combines several individual
characteristics is the second order autocorrelation $g^{(2)}$ which relates the single-photon and
multi-photon rates for a source. The best-performing single-photon sources in terms of this parameter, and hence for
a number of important applications, are heralded single-photon sources (HSPS's) \cite{HSPS1,
HSPS2}, and not single-quantum-emitter-based sources. These sources rely on photons produced in
pairs, with the detection of one photon heralding the existence of the other. While heralded
sources are probabilistic, because the creation of a photon pair is intrinsically probabilistic,
rather than deterministic,  the ability to herald the output photon offers significant advantage
and utility  \cite{shell_2, alan}. The heralded photon output of these sources, however, often
suffers from background photons that pollute the output and can contribute to noise in a particular
application. But here again,  the ability to herald the output photon can in many cases be used to
mitigate the presence of background photons in the output channel by allowing for filtering based
on temporal post-selection of measurement events \cite{NHSPS}.

There are however, some cases where temporal post-selection cannot be used, for example when
detectors have low temporal resolution (as e.g. in Ref. \cite{noiTes}). A useful alternative is an
HSPS coupled with an optical shutter that reduces the presence of unwanted photons in the heralded
channel \cite{Rarity, NHSPS, Lobino}. Such a combination takes this source much closer to a realm
of true single-photon sources, where, while a heralded pair source shows the antibunching required
for a single-photon source only when conditioned on the herald photon, this source exhibits
antibunching even when operating without regard to the heralding.

Here we present the implementation of such a low-noise HSPS with an extremely low level of
background, that does not need temporal post-selection of the heralded counts. For this reason and
for many applications, it has advantages over sources based on single quantum emitters, offering
much better performance characteristics in terms of both the multi-photon component, characterized
by $g^{(2)}(0)=0.005(7)$, and the noise  characterized by the \textit{Output Noise Factor} (ONF),
defined as the ratio of the background photons to the total photons counted by the two detectors
\footnote{$ONF$, introduced in \cite{NHSPS}, is defined as \\$
    ONF=\frac{N_{1}^{\mathrm{(Bkg)}}+N_{2}^\mathrm{(Bkg)}}{N_{1}^{\mathrm{(True)}}+N_{1}^{\mathrm{(Bkg)}}+
    N_{2}^{\mathrm{(True)}}+N_{2}^{\mathrm{(Bkg)}}
    }$, where $N_{i}^{\mathrm{(Bkg/True)}}$
    corresponds to the number of background/true counts measured by the $i^{\textrm{th}}$ detector.  },
$ONF=0.25(1)\%$ (all uncertainties are 1-$\sigma$ standard deviations). Note that $g^{(2)}(0)$ is
measured in a Hanbury-Brown Twiss (HBT) setup with two non photon-number resolving detectors. Such
measurements with relatively low detection efficiency and/or for antibunched sources allow the
ratio of coincidences to the product of the singles (often denoted by $\alpha$ \cite{grangier1}) to
be a good measure of $g^{(2)}(0)$. Only one single photon source has been demonstrated with a
comparable $g^{(2)}(0)$ \cite{giacobino}, but that value was estimated after background subtraction
(without which, the result was 10x higher).

\begin{figure}
\begin{center}
\includegraphics[width=0.90\textwidth]{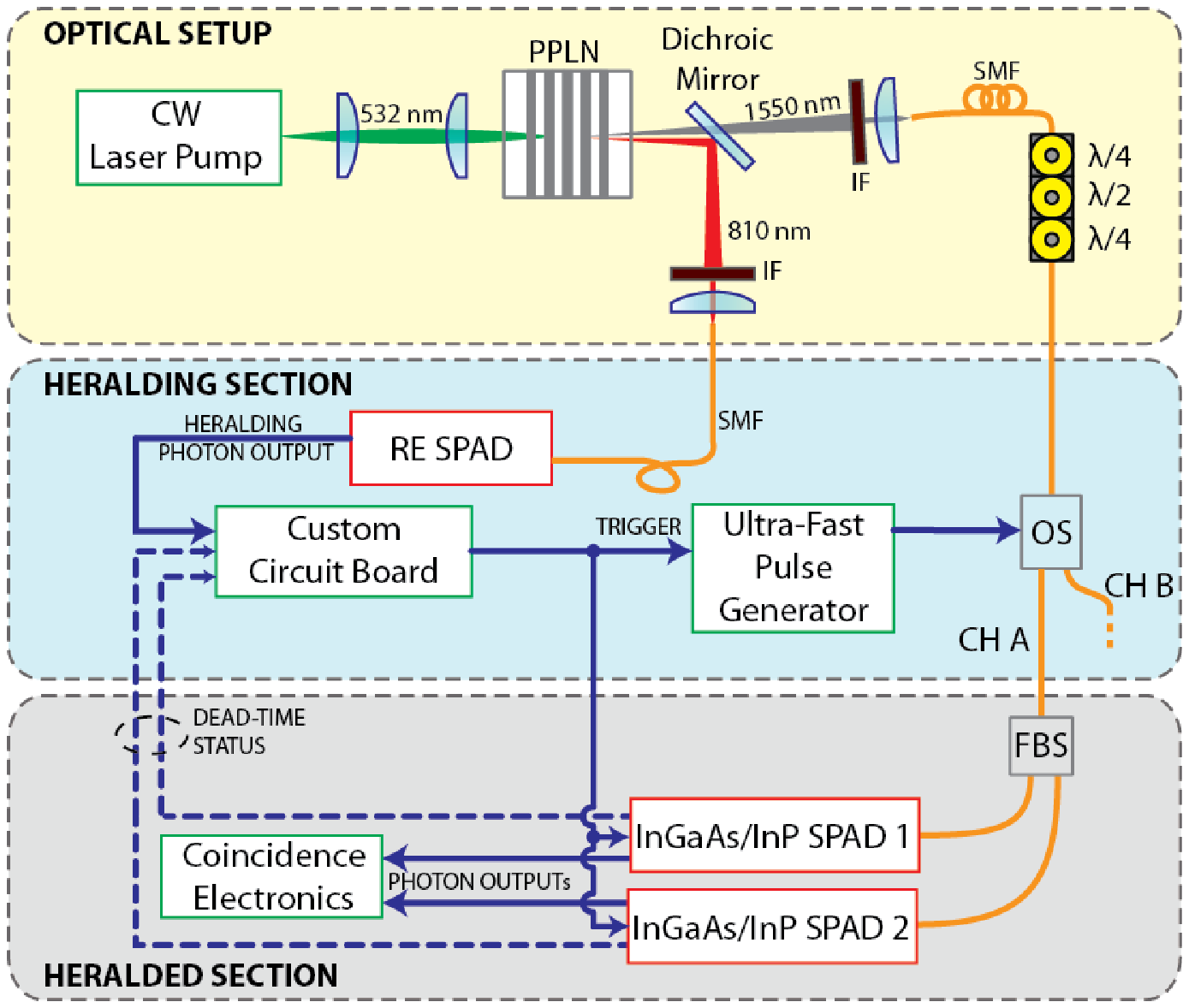}
\end{center}
\caption{ Experimental setup. \label{f:setups}}
\end{figure}

Our setup (Fig.\ref{f:setups}) is comprised of a 532 nm continuous wave (CW) laser that pumps a
10 mm $\times$ 1 mm $\times$ 10 mm periodically-poled lithium niobate (PPLN) crystal. The crystal generates
photon pairs at $\lambda_{\textrm{s}}=1550$ nm (signal) and $\lambda_{\textrm{i}}=810$ nm (idler) respectively via
non-degenerate parametric down-conversion. The idler photon passes through an interference
filter (IF) and through an iris (to reduce the source bandwidth below 10~nm
full-width-at-half-maximum (FWHM)), and then is injected into a single-mode fiber (SMF) connected
to a heralding detector based a silicon-based single-photon avalanche diode (SPAD). This optical
configuration is similar to previous experiments, thus the coupling efficiency of the heralded
photon was  $\approx13\%$\cite{NHSPS, metro}.

Our SPAD is a prototype module based on a Red-Enhanced SPAD \cite{MPD, GulinattiJMO,disclaimer}
designed to combine a high photon detection efficiency of $ \approx 40 \%$ in the near infrared
region with a low timing jitter of $ \approx 90$ ps FWHM at 810 nm, the wavelength of the heralding
photons in this experiment. Both of these parameters have a strong impact on system performance. A
lower detection efficiency would result in the loss of a significant number of heralding photons
and therefore of the corresponding heralded counterparts. On the other hand, low timing jitter is
critical to precisely discriminate the arrival time of heralded photons, thereby allowing for
effective background suppression.

The heralded signal is spectrally selected by an iris and an IF (corresponding to a bandwidth of
less than 30~nm FWHM), and it is sent through 20~m of SMF fiber, connected to a commercial
ultra-high-speed 2x2 optical switch (OS) based on a  LiNbO$_3$ integrated waveguide Mach-Zehnder
interferometer \cite{EOSpace}. The OS is operated by a custom-made circuit, with a timing jitter
below 6~ps FWHM, that receives the heralding signal from the heralding SPAD and triggers a
custom-made fast pulse generator (with 50 ps rise-time) for switching from channel B to channel A
for a time interval $\mathcal{T}_{\textrm{open}} \approx $ few~ns.

The output of our source, i.e. Channel A of the optical switch (Fig. \ref{f:setups}), is sent to  a
HBT interferometer for characterization. The HBT interferometer consists of  a $50:50$ fiber beam
splitter (FBS) whose outputs are connected to two InGaAs/InP SPADs with high detection efficiency ($30
\%$ at 1550 nm), low timing jitter (160 ps FWHM), and low dark count rate ($\approx$~20~kcps)
\cite{Tosirevsci}.

The outputs of the two InGaAs/InP SPADs are sent to coincidence electronics with a time-bin
resolution of 2~ps and to a multi-channel analyzer whose output is collected by a computer. To
reduce spurious detections and to increase the signal-to-noise ratio, the InGaAs/InP SPADs are
operated in a gated mode, i.e. they are turned on only when a heralding photon is detected.

For each accepted heralding count, the custom control circuit triggers a pulse generator that opens
the OS for a chosen time interval $\mathcal{T}_{\textrm{open}}$; a SPAD count is considered a valid
herald by the control circuit only when both the InGaAs/InP SPADs have recovered from any previous
dead times and are ready to count and incoming photon. To lower the probability of afterpulsing,
the dead time for the two InGaAs detectors is set to 50~$\mu$s.

\begin{figure}
\begin{center}
\includegraphics[width=0.90\textwidth]{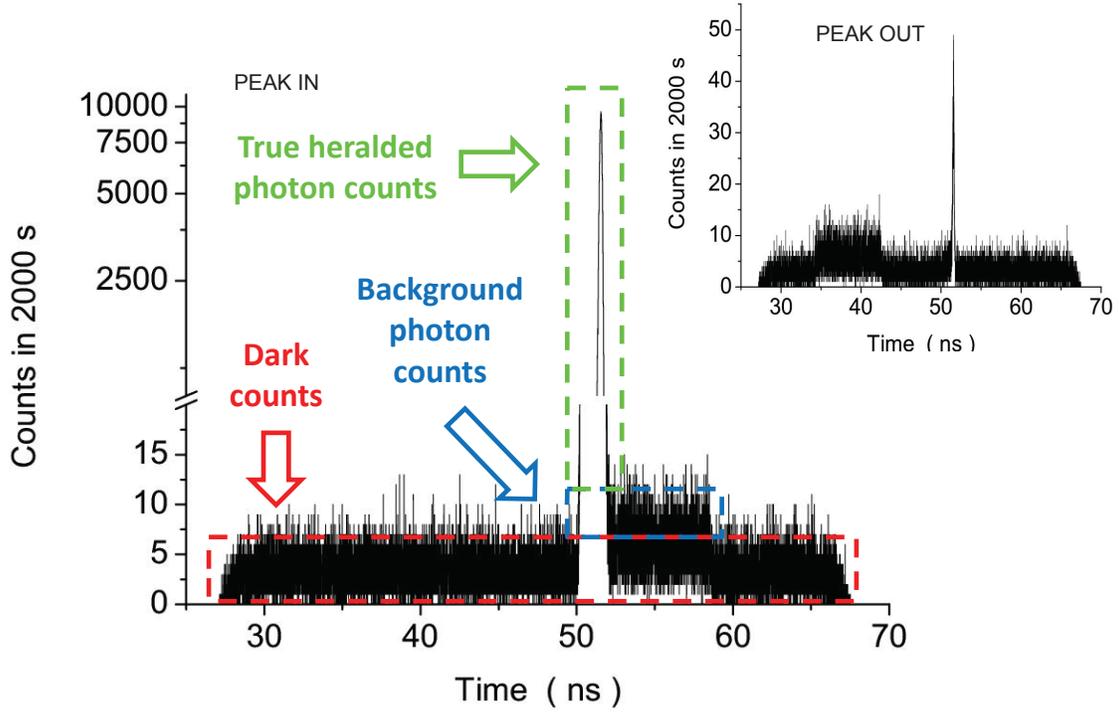}
\caption{ Histograms of counts from SPAD1, which is gated on for 40 ns. The OS of the HSPS is
opened for $\mathcal{T}_{\textrm{open}}=10$ ns). Main figure: the heralded photon peak timed to
arrive during switch open time (i.e. Channel A opening is properly synchronized with the incoming
heralded photons); true, background, and dark count contributions are clearly seen. Inset: the OS
open time is set to miss the heralded photon peak, thus it is highly suppressed: the ratio of the
integrals of the heralded photon peaks in these two configurations is
$\approx10^{-3}$.}\label{peak-in}
\end{center}
\end{figure}

To study the behavior of our source with respect to the duration of the OS open time
$\mathcal{T}_{\textrm{open}}$, we made measurements at five values of $\mathcal{T}_{\textrm{open}}$
(20~ns, 16~ns, 10~ns, 5~ns, and 2~ns). A histogram of counts from SPAD1, gated on for 40~ns by the heralding counts (Fig. \ref{peak-in})
shows three distinct features: one due to true heralded photon counts, one due to background and
stray light photons emitted while the OS is open, and another due to detector dark counts.

Among the possible figures of merit \cite{shell_2, alan,met,onoff} than can be used for source
characterization, the \textit{ONF} introduced in \cite{NHSPS} and the $g^{(2)}(0)$ are the most
appropriate for the source here with its extremely low noise level and its low multiphoton content.

\begin{figure}
\begin{center}
\includegraphics[width=0.90\textwidth, angle=0]{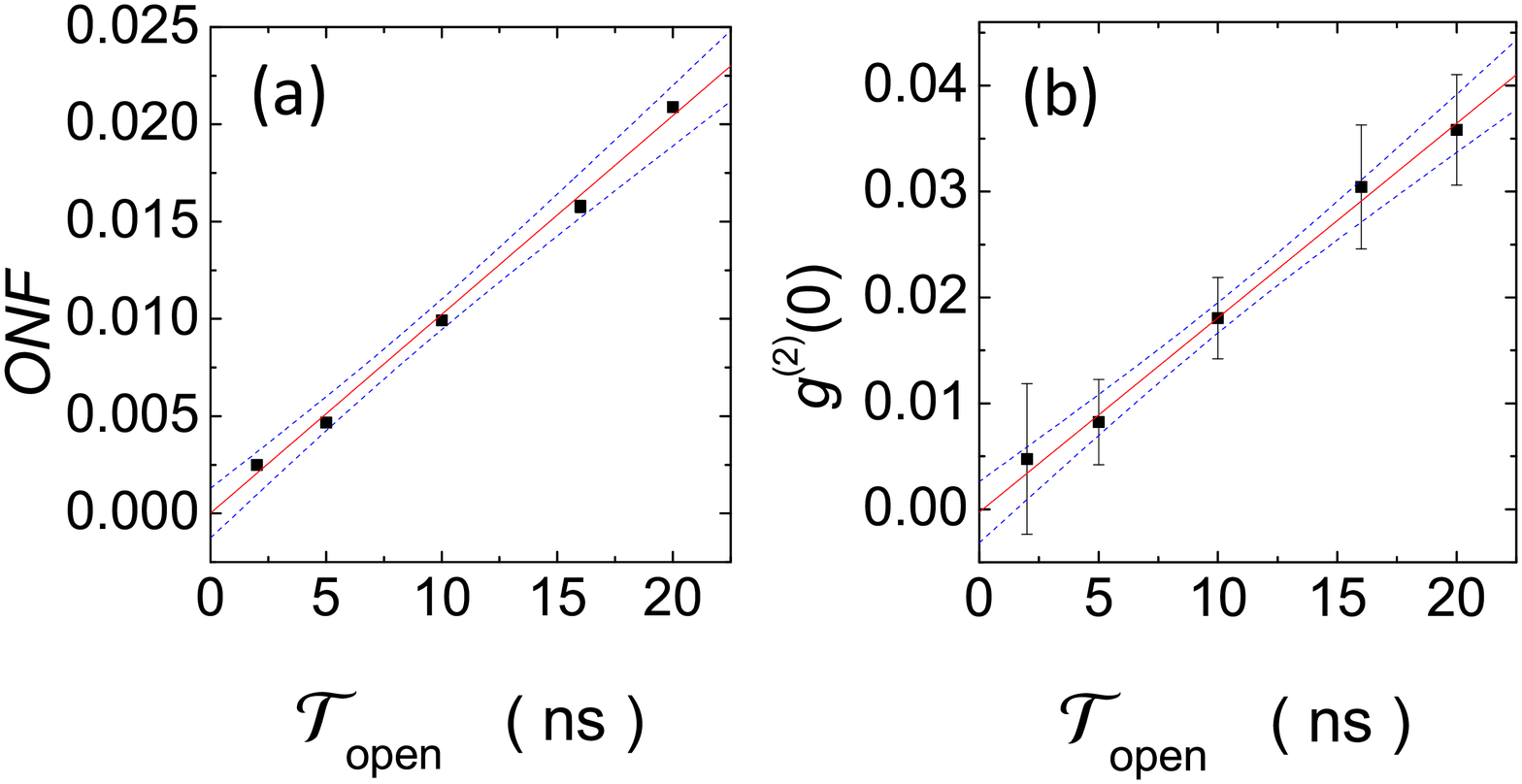}
\vspace{1cm}
  \caption  {(a) $ONF$ and (b) $g^{(2)}(0)$ versus the switching time $\mathcal{T}_{\textrm{open}}$. The linear fits
(solid lines) of the data (points) are shown, together with 95\% confidence bands (dashed curves).}
\label{ONF-plot}
\end{center}
\end{figure}

In Fig. \ref{ONF-plot} (a), we see that the $ONF$ decreases linearly with decreasing
$\mathcal{T}_{\textrm{open}}$, a direct consequence of fewer background photons emitted as the OS
is open for less time. The minimum $ONF$ value achieved with this current setup, at
$\mathcal{T}_{\textrm{open}}=2$ ns, is $0.25(1)\%$ (almost six times lower than the $1.4\%$
reported in \cite{NHSPS}), showing the great improvement achieved with the use of low-jitter SPADs
and fast electronics
\footnote{A further reduction of $\mathcal{T}_{\textrm{open}}$ would obviously reduce $ONF$ and
$\alpha$ values, but would also lose true heralded photons. }.

The same linear dependance with $\mathcal{T}_{\textrm{open}}$ is seen for $g^{(2)}(0)$, in Fig.
\ref{ONF-plot} (b): in this case, the lowest measured value  is $0.005(7)$. This
is the lowest ever obtained for a single-photon source. We compare it with the previous best
achievements for single-photon sources that are not based on postselection (i.e. heralding)
\cite{alan}, for example, to the $g^{(2)}(0)$ obtained for $^{40}$Ca$^{+}$ in an ion-trap cavity
($g^{(2)}(0) = 0.015 $) \cite{sps2}, or to a quantum dot in micropillar ($g^{(2)}(0) = 0.02$)
\cite{sps1_1,sps1_2} or, to the already mentioned result of Ref. \cite{giacobino} where it was
evaluated to be $g^{(2)}(0)= 0.004$ but only after background subtraction, leaving our source with
the lowest two-photon component.

Furthermore, the linear fit to $g^{(2)}(0)$ (Fig.~\ref{ONF-plot}(b)), yields an ideal zero-noise
configuration ($\mathcal{T}_{\textrm{open}}= 0$) of $g^{(2)}(0) = -0.0003(9)$, whose compatibility
with 0 shows that there is no measurable two-photon content in the true heralded peak, making ours
an ultra-pure single-photon source. Analogous considerations hold for the $ONF$ data in Fig.~\ref{ONF-plot}(a).
The fit shows that $ ONF = 0.00002(40)$
for $\mathcal{T}_{\textrm{open}}= 0$, i.e. consistent with zero.

The relative uncertainties on the $g^{(2)}(0)$ values are quite large with respect to those
obtained for the $ONF$. This is because the double-detection events needed to evaluate $g^{(2)}(0)$
are relatively rare, further highlighting the extremely low noise of our HSPS.

To check the extinction performance of our OS, we compared the true heralded-photons in the
\textit{peak-in} and \textit{peak-out} configurations (see Fig. \ref{peak-in}): the ratio between
the integrals of these two peaks (once the background and dark counts have been subtracted) defines
the OS extinction factor $r$ \cite{NHSPS}. Our measurements yielded a
value of $r=0.0010(2)$   
(compatible with the factory specifications of the switch), and more than three times better than
the one obtained in \cite{NHSPS}. This was possible by improved polarization control, as well as a
better spectral selection of the heralded photons, reducing their bandwidth and making our
interferometric optical shutter work at its best.

In conclusion, through the use of high- detection-efficiency low-jitter SPADs and fast electronics
custom-made for operating an optical switch, we have realized a high performance heralded
single-photon source in terms of single-photon purity
$g^{(2)}(0)$ and $ONF$. 
The minimum $g^{(2)}(0)$ value observed, $0.005$, surpasses the best reported single-photon
sources, including those based on single emitters \cite{shell_2, alan,met}.

Furthermore, with respect to other single-photon sources like nitrogen vacancy centers in nanodiamonds, quantum
dots, etc. \cite{shell_2, alan}, this low-noise HSPS has the advantage of the wide wavelength
tunability typical of parametric-down-conversion-based sources, and, since our heralded photons are in the telecom band,
this source is compatible with available telecom components. Its special features, like the
elimination of background photons and the ability to choose a dead time $t_{\textrm{dead}}$ in
which the control circuit ignores the incoming heralds, can be very useful when dealing with slow
response systems and detectors (e.g. the transition-edge superconducting microbolometers
\cite{TES}), where the slow response does not allow temporal discrimination of unwanted events.

Finally, we note that the performance of this source can be further enhanced by integrating the
whole system in a single PPLN crystal, the same technology used for the fast OS exploited here as
an optical shutter.

\textit{Acknowledgements.} The research leading to these results has received funding from the
European Union on the basis of Decision No. 912/2009/EC (project IND06-MIQC), by MIUR, FIRB
RBFR10YQ3H and RBFR10UAUV, and by Compagnia di San Paolo.

\newpage

\newpage

\newpage

\begin{thebibliography}{99}
\bibitem{disclaimer} Certain commercial equipment, instruments or materials are identified in this paper to foster
understanding. Such identification does not imply recommendation or endorsement by the National
Institute of Standards and Technology, nor does it imply that the materials or equipment are
necessarily the best available for the purpose.
\end{thebibliography}
\end{document}